# Antiferromagnetic Ordering Enhanced Magnetic Damping in $Mn_2Au$/CoFeB Bilayers

Donghang Xie[1,2] , Haozhe Wang[1,3] , Zhe Zhang[1,2] , Zishuang Li[1,3] , Jiahua Lu[1,2] , Ronghua Liu[1,3,a)] , Jun Du[1,3], Bo Liu[1,2] , Yu Yan[1,2] , Liang He[1,2] , Jing Wu[4] , Rong Zhang[2] , Bo Liu[5], Tiejun Zhou[5], Yongbing Xu[1,2,4,a)] , Xuezhong Ruan[1,2,a)]

## AFFLIATIONS

*1 State Key Laboratory of Spintronics, Nanjing University, Suzhou, 215163, China*

*2 Jiangsu Provincial Key Laboratory of Advanced Photonic and Electronic Materials, School of Electronic Science and Engineering, Nanjing University, Nanjing 210093, China*

*3 School of Physics, Nanjing University, Nanjing 210093, China*

*4 York-Nanjing Joint Center (YNJC) for Spintronics and Nano Engineering, Department of Electronics and Physics, The University of York, York YO10 5DD, UK*

*5 State Key Laboratory for Spintronic Devices and Technologies, Hangzhou 311305, China*

[a)] Authors to whom correspondence should be addressed: rhliu@nju.edu.cn, ybxu@nju.edu, and xzruan@nju.edu.cn.

## ABSTRACT

Antiferromagnets (AFMs) hold significant potential for spintronic devices owing to their insensitivity to external magnetic fields and the absence of stray fields. Beyond these inherent advantages, an AFM can manipulate the magnetic dynamics of a ferromagnet (FM) layer in AFM/FM bilayers, whereas the mechanism of such manipulation remains controversial. Here, we investigate the magnetic dynamics of AFM/FM $Mn_2Au$/CoFeB bilayers via Ferromagnetic Resonance (FMR). It is found that the Néel temperature of 2-nm-thick $Mn_2Au$ is as low as ~40 K, in sharp contrast to that of bulk $Mn_2Au$, which exceeds 1000 K. In the $Mn_2Au$(2 nm)/CoFeB(4 nm) bilayer, the magnetic damping $\alpha$ of the CoFeB layer increases from 0.013 to 0.047 as temperature decreases from 160 K to 10 K, accompanied by a synchronous increase in the exchange coupling field $H_{rot}$. Such an increase in $\alpha$ is attributed to the enhanced spin angular momentum transfer from CoFeB to $Mn_2Au$, mediated through AFM-FM exchange coupling between $Mn_2Au$ and CoFeB, which is enhanced by the $Mn_2Au$ antiferromagnetic ordering as the temperature decreases. Our study provides deeper insights into AFM/FM dynamics and spintronic storage technology.

In recent years, Magnetic Random Access Memory (MRAM) has attracted increasing attentions in the field of modern storage technology due to its low power consumption, high speed, and non-volatility, positioning it as a promising alternative to conventional storage devices.[1,2] The operation of MRAM relies on the switching of magnetic moments between the “0” and “1” states in magnetic tunnel junctions (MTJs) such as CoFeB/MgO/CoFeB.[3,4] Therefore, understanding and controlling the magnetization dynamics of these ferromagnets (FMs) has become a central focus in both fundamental research and industrial applications. In ferromagnetic dynamics, the magnetic damping and the anisotropy field are two key parameters that determine the critical switching current density and the stable orientation of magnetic moments, respectively.[5] Consequently, exploring effective approaches to modulate these parameters is essential for the development of magnetic storage technology.

Antiferromagnetic (AFM) materials, insensitive to external magnetic fields and free of stray fields,[6–10] hold significant promise for next-generation magnetic storage devices. Beyond these intrinsic features, the antiferromagnet/ferromagnet (AFM/FM) exchange-coupled systems exhibit intriguing properties, including improved thermal stability,[11–13] achievement of perpendicular magnetic anisotropy (PMA) in the FM layer,[14–18] and field-free THz emission.[19] Moreover, the magnetic dynamics of the FM layer can be significantly modulated by the adjacent AFM. Fan *et al.* studied the magnetic dynamics in CoO/Fe and attributed the enhanced damping to exchange coupling.[20] Zhang *et al.* correlated the increased damping in FeGa/IrMn with the ultrafast demagnetization rates and attributed it to the competition between FM and AFM spin pumping.[21] Frangou *et al.* emphasized spin pumping as the dominant mechanism responsible for the magnetic dynamics in AFM/FM bilayers.[22] Liu *et al.* proposed that the interfacial exchange spring effect is responsible for the damping modulation in IrMn/CoFe.[23] Despite these reports, no consensus exists on the mechanism by which the AFM manipulates the magnetic dynamics of FM. These aspects highlight the need for further investigation into the magnetic dynamics of AFM/FM heterostructures.

Furthermore, investigations on the modulation of magnetic dynamics in FM by an adjacent AFM layer have motivated the exploration of novel AFM materials. As a member of the collinear AFM family, $Mn_2Au$ exhibits a robust AFM-FM exchange coupling effect[24,25] and strong sublattice magnetization (~$4\mu_B$).[26] Its Néel temperature ($T_N$) exceeds 1000 K, endowing $Mn_2Au$ with excellent thermal stability for high-temperature spintronic applications.[26,27] Although theory predicts that the $T_N$ of an AFM film may be suppressed when its thickness is reduced to a few nanometers,[28,29] it remains unclear whether the AFM order in $Mn_2Au$ persists at such ultrathin thicknesses. If it does, how would an ultrathin $Mn_2Au$ layer affect the magnetic dynamics of an adjacent FM layer in the $Mn_2Au$/FM bilayers? Therefore, in this work, we employ temperature-dependent ferromagnetic resonance (FMR) to systematically investigate the magnetic dynamics of $Mn_2Au$/CoFeB bilayers with nanometer-thick $Mn_2Au$ layers. Superconducting Quantum Interference Device (SQUID) measurements confirm the AFM order in $Mn_2Au$ with a thickness down to 2 nm. Temperature-dependent FMR measurements on $Mn_2Au$(2 nm)/CoFeB bilayer show that magnetic damping ($\alpha$) remains nearly unchanged and the exchange coupling field ($H_{rot}$) is absent above 160 K, while both $\alpha$ and $H_{rot}$ increase synchronously as temperature decreases below 160 K. In contrast, for the reference sample CoFeB, the $\alpha$ remains nearly unchanged and $H_{rot}$ is absent from room temperature to 10 K. The damping increase is eliminated by inserting an Al interlayer between $Mn_2Au$ and CoFeB layers, ruling out the spin pumping effect, and indicating that the enhanced damping is due to the enhanced spin angular momentum transfer mediated by the AFM-FM exchange coupling, which is strengthened by the $Mn_2Au$ antiferromagnetic ordering as the temperature decreases. Our study provides deeper insight into the mechanism of the AFM-FM exchange-coupled systems and their potential in magnetic storage applications.

The following structures were deposited on Si substrates by Magnetron Sputtering: Si/MgO(3)/$Mn_2Au$(2)/$Co_{40}Fe_{40}B_{20}$(4)/Al(3) (denoted as $Mn_2Au$(2)/CoFeB), Si/MgO(3)/$Mn_2Au$(4)/$Co_{40}Fe_{40}B_{20}$(4)/Al(3) (denoted as $Mn_2Au$(4)/CoFeB), Si/MgO(3)/$Mn_2Au$(2)/Al(1.5)/$Co_{40}Fe_{40}B_{20}$(4)/Al(3) (denoted as $Mn_2Au$(2)/Al/CoFeB) and the reference sample Si/MgO(3)/$Co_{40}Fe_{40}B_{20}$(4)/Al(3) (denoted as CoFeB). The numbers in the parentheses are the layer thickness in nanometers. The MgO seed layer was deposited by RF sputtering to optimize the growth of $Mn_2Au$ and CoFeB. All other layers were deposited by DC sputtering. $Mn_2Au$ was sputtered from an alloy target (atomic ratio: Mn: Au=2:1). The base pressure was better than $9\times10^{-8}$ Torr, and the working argon pressure was set at 7 mTorr. The Al capping layer prevented oxidation. Static magnetic properties were characterized using a Superconducting Quantum Interference Device (SQUID). The magnetic dynamics were studied by Ferromagnetic Resonance (FMR). In the FMR setup shown in Fig. 2(a), the sample was placed face-down on a coplanar waveguide (CPW), ensuring optimal contact with the radio-frequency field $H_{rf}$, generated by an rf-current $I_{rf}$ applied through the CPW at various frequencies. Measurements were conducted in a vacuum (better than $1\times10^{-5}$ Pa) from room temperature (RT) to 10 K. A second coil was installed in the magnet, and a lock-in amplifier (LIA) was used to improve the signal quality.

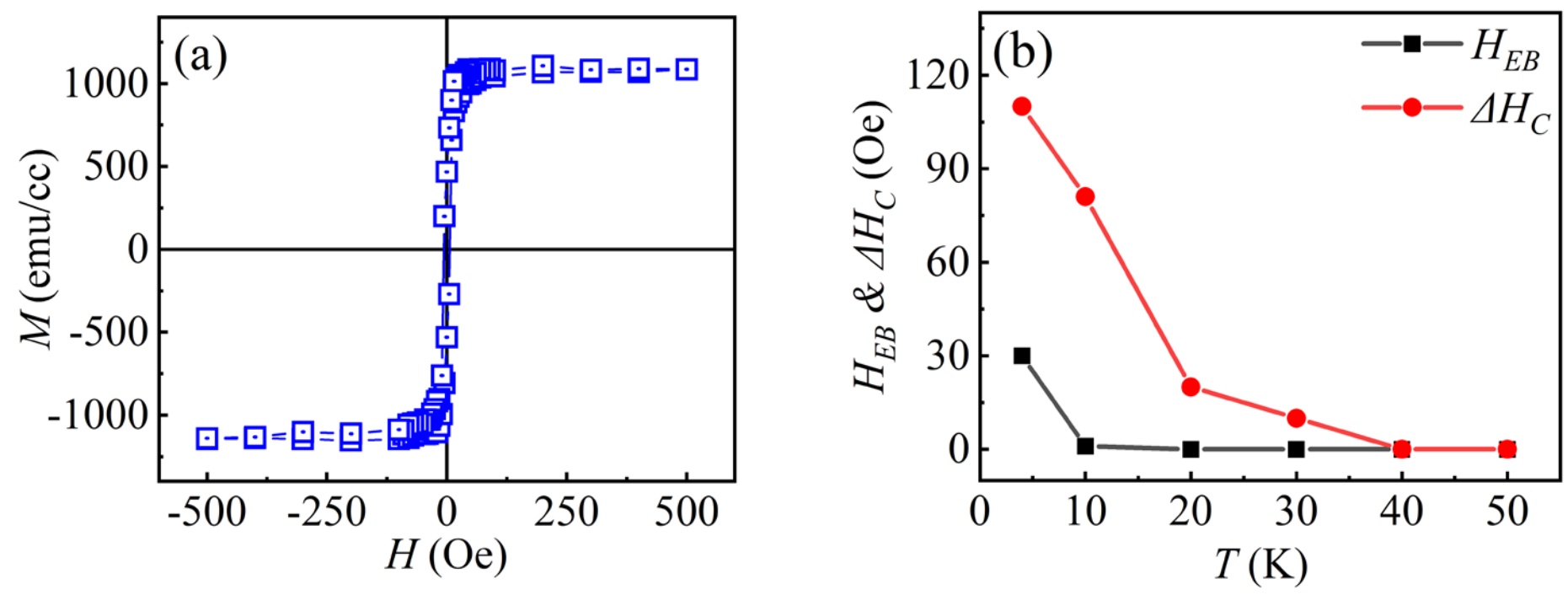

**FIG. 1.** (a) In-plane magnetic hysteresis loop of $Mn_2Au$(2)/CoFeB at RT. The saturation magnetization $M_S$ is 1140 emu/cc. (b) Temperature-dependent exchange bias field $H_{EB}$ and increment of coercivity $\Delta H_C$.

The magnetic hysteresis loop of $Mn_2Au$(2)/CoFeB at RT is presented in Fig. 1(a). The saturation magnetization $M_S$ is 1140 emu/cc, consistent with reported values for CoFeB.[30–33] To verify AFM order, field-cooled magnetic hysteresis loops were measured using SQUID (see supplementary material, Fig. S1). The sample was cooled from RT to 4 K under an in-plane field of 5 kOe. Exchange bias field $H_{EB}$ and increment of coercivity $\Delta H_C$ ($\Delta H_C = H_C - H_C(RT)$) were observed at 4 K, confirming the presence of macroscopic AFM order in the 2 nm $Mn_2Au$ layer at low temperature. As the temperature increases, $H_{EB}$ disappears at around 10 K, and $\Delta H_C$ disappears at around 40 K (see supplementary material, Fig. S1 and Fig. 1(b)). In contrast, the coercivity of a stand-alone CoFeB layer remains unchanged between RT and 5 K.[4] As reported by Lenz *et al.*, the enhanced coercivity in AFM/FM bilayer arises from AFM order, and the temperature at which $\Delta H_C$ disappears upon warming is regarded as the Néel temperature $T_N$.[18,34] Accordingly, we conclude that the $T_N$ of 2 nm $Mn_2Au$ is approximately 40 K.

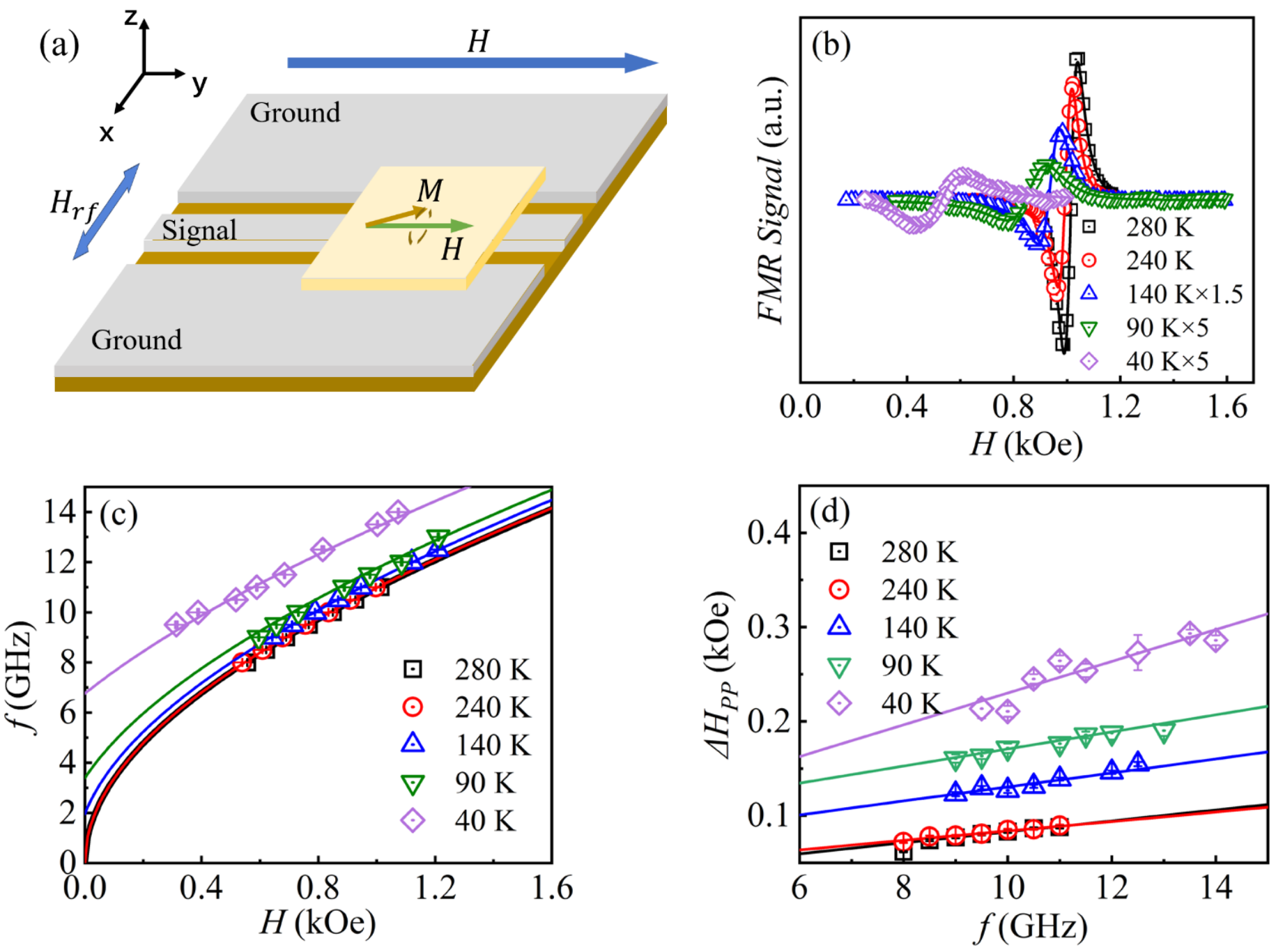

**FIG. 2.** (a) Schematic illustration of the FMR setup. (b) FMR signals at 11 GHz $H_{rf}$ for $Mn_2Au$(2)/CoFeB at 280K, 240K, 140K, 90K and 40 K. The solid lines are the fitting results of Equation (1). (c) FMR dispersion relation ($f - H_{res}$). (d) Frequency-dependent FMR linewidth ($\Delta H_{PP} - f$). The linewidths at 40 K and 90 K are shifted downward by 0.05 kOe for a better comparison. The solid lines are fits to Equation (2) and (3), respectively.

Dynamic magnetic properties were characterized via ferromagnetic resonance (FMR). Fig. 2(a) presents the schematic of the FMR setup. A radio-frequency magnetic field ($H_{rf}$) with variable frequency ($f$) was applied along the x-axis, while a static magnetic field ($H$) was applied along the y-axis. By sweeping $H$, the magnetization of the FM layer resonates when $H$ matches the resonant field ($H_{res}$) for a given frequency of $H_{rf}$, leading to absorption of the rf field, detected as the FMR signal. Fig. 2(b) presents the experimental FMR signals from $Mn_2Au$(2)/CoFeB as a function of the magnetic field, measured at a fixed $H_{rf}$ frequency of 11 GHz and at various temperatures. These signals are fitted by the derivative of a Lorentzian function:[35]

$$y = y_0 - S\frac{2\cdot\Delta H^2\cdot(H-H_{res})}{[(H-H_{res})^2+\Delta H^2]^2} + A\frac{[\Delta H^2-(H-H_{res})^2]\cdot\Delta H}{[(H-H_{res})^2+\Delta H^2]^2} \quad (1)$$

In this equation, the extracted $\Delta H$ is half of the peak-to-peak linewidth $\Delta H_{PP}$ and $H_{res}$ is the resonant field. As shown in Fig. 2(b), two prominent features emerge in the FMR signal as the temperature decreases. First, the resonance field $H_{res}$ systematically shifts toward zero. Second, the peak-to-peak linewidth $\Delta H_{PP}$ of the FMR signals broadens significantly. Here, the shift of $H_{res}$ suggests an *"internal"* magnetic field, which is built up inside the sample and effectively compensates the applied external field. Furthermore, such a shift is isotropic in the film plane, as indicated by the isotropic in-plane angular-dependence of $H_{res}$, as shown in the supplementary material Fig. S2. This behavior resembles the rotatable anisotropic field ($H_{rot}$) originated from the interfacial exchange coupling between unpinned uncompensated AFM spins and FM spins in AFM/FM heterostructures.[36–38] Therefore, we also refer to $H_{rot}$ as the "exchange coupling field" in this work. The dispersion relation ($H_{res}$ as a function of resonant frequency $f$) is plotted in Fig. 2(c) and is well described by the following equation:[39]

$$f = \gamma\sqrt{(H_{res} + H_{rot} + H_{EB})(H_{res} + H_{rot} + H_{EB} + 4\pi M_{eff})} \quad (2)$$

where $\gamma$, $4\pi M_{eff}$, $H_{rot}$, and $H_{EB}$ are the gyromagnetic ratio, effective demagnetization field, rotatable anisotropy field, and exchange bias field, respectively. We used 2.8 GHz/kOe for $\gamma$ in our fitting.[40] The value of $H_{EB}$ was obtained from the shift of the magnetic hysteresis loop measured by SQUID (see Fig. 1(b)). Here, $4\pi M_{eff} = 4\pi M_S - \frac{2K_\perp}{M_S}$, where $K_\perp$ is the perpendicular anisotropy constant. For $Mn_2Au$(2)/CoFeB, the $4\pi M_{eff}$ was determined to be 14.5 kOe from the saturation field of the out-of-plane hysteresis loop measured at RT (see supplementary material, Fig. S3). This value is close to the demagnetization field $4\pi M_S$=14.3 kOe. The close agreement between $4\pi M_{eff}$ and $4\pi M_S$ indicates that $K_\perp$ is negligible. As temperature decreases, the increase in $M_S$ renders the term $\frac{2K_\perp}{M_S}$ even more negligible. Therefore, in our analysis, we approximated $4\pi M_{eff}$ as $4\pi M_S$, with $4\pi M_S$ obtained from the *M-T* curve measured by SQUID (see supplementary material, Fig. S4).

The extracted peak-to-peak linewidth $\Delta H_{PP}$ as a function of $f$ is shown in Fig. 2(d). The magnetic damping is obtained by fitting the data with the following equation:[41]

$$\Delta H_{PP} = \Delta H_0 + \frac{2}{\sqrt{3}}\frac{\alpha}{\gamma}f \quad (3)$$

Here, $\Delta H_0$ denotes the inhomogeneous linewidth broadening and $\alpha$ is the effective magnetic damping. As shown in Fig. 3(a), the effective magnetic damping of the reference sample CoFeB remains nearly invariant at ~0.011 from RT down to 10 K. This invariability is consistent with the reported results.[42] However, when the CoFeB layer is deposited on 2-nm-thick $Mn_2Au$, the temperature dependence of damping changes dramatically. From RT down to ~160 K, the magnetic damping of $Mn_2Au$(2)/CoFeB remains nearly unchanged at about 0.013. As the temperature further decreases below 160 K, the magnetic damping increases pronouncedly, reaching 0.047 at 10 K. Meanwhile, Fig. 3(b) shows the temperature dependence of the extracted rotatable anisotropic field $H_{rot}$. For the reference sample CoFeB, $H_{rot}$ is nearly zero over the entire temperature range. Thus, $H_{rot}$ of the reference sample CoFeB is negligible. In contrast, the $H_{rot}$ of $Mn_2Au$(2)/CoFeB remains near 0 kOe between 160 K and RT, and increases dramatically below about 160 K. This temperature dependence is analogous to that of $\alpha$ for $Mn_2Au$(2)/CoFeB. The synchronous increase of $\alpha$ and $H_{rot}$ with decreasing temperature indicates that they should share a common physical origin in the $Mn_2Au$/CoFeB bilayer.

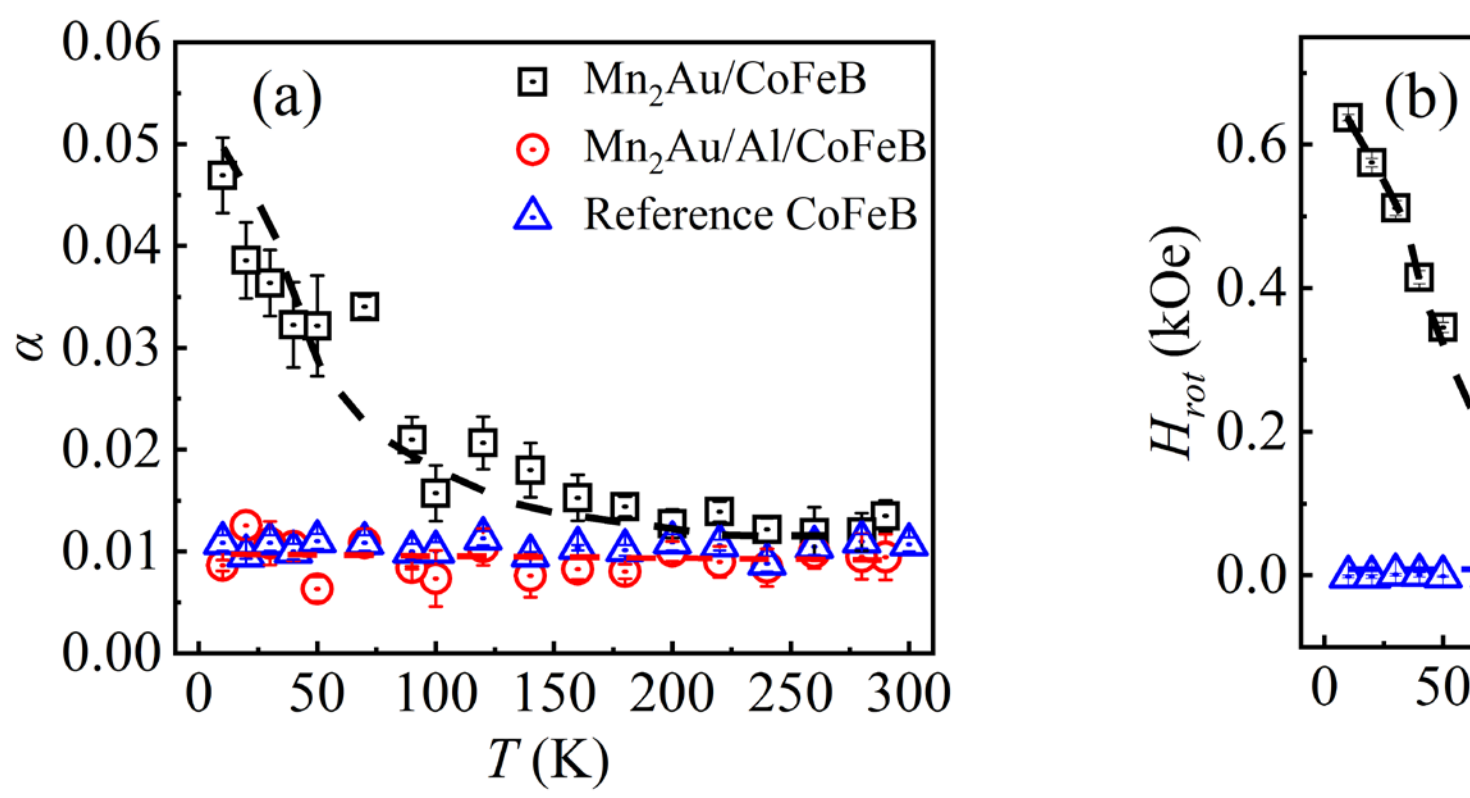

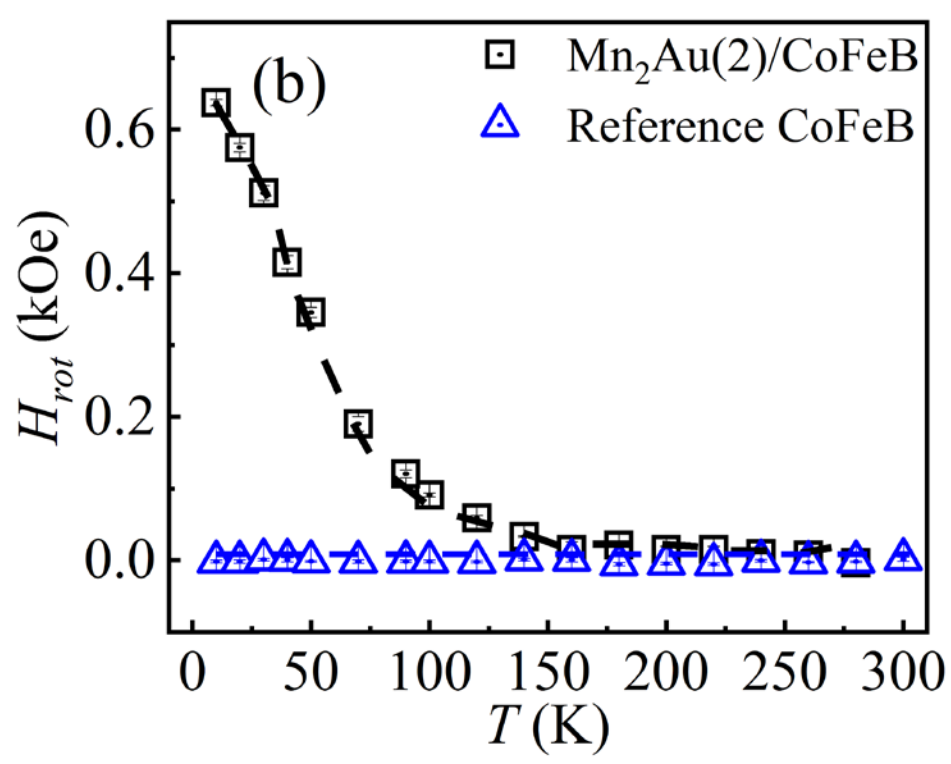

**FIG. 3.** (a) Temperature-dependent magnetic damping $\alpha$ of $Mn_2Au(2)/CoFeB$, $Mn_2Au(2)/Al(1.5)/CoFeB$ and reference sample CoFeB. (b) Temperature-dependent rotatable anisotropic field $H_{rot}$ of $Mn_2Au(2)/CoFeB$ and reference sample CoFeB. The dashed lines are the visual guides.

The temperature dependence of $\alpha$ and $H_{rot}$ in the $Mn_2Au(2)/CoFeB$ bilayer contrasts markedly with that of the reference sample CoFeB. Two-magnon scattering (TMS)—a process in which a magnon with zero wave-vector ($k$=0) scatters into degenerate, nonuniform spin waves ($k$≠0) due to fluctuating exchange coupling and exchange bias at the interface[21]—has been widely studied in AFM/FM and heavy metal/FM bilayers. Previous studies have shown that TMS can be suppressed by an externally applied field or exchange coupling field, leading to a decrease in magnetic damping.[20,32] This feature contradicts our results in Fig. 3, where the $\alpha$ increases as the exchange coupling field $H_{rot}$ increases in $Mn_2Au(2)/CoFeB$. This discrepancy indicates that the TMS is not the dominant mechanism in $Mn_2Au/CoFeB$. The good linearity in the FMR linewidth vs. frequency curves in Fig. 2(d) also indicates that TMS is negligible.[43,44] Moreover, the polar-angle-dependent linewidth further excludes the contribution of TMS, as shown in Fig.S5 and discussed in the supplemental materials.

The spin pumping effect is known to modify magnetic damping in AFM/FM or heavy-metal/FM heterostructures by transporting electronic spin current.[21,45,46] To investigate whether the enhanced damping below 160 K arises from spin pumping, we inserted a 1.5 nm Al interlayer between the $Mn_2Au$ and CoFeB layers. Aluminum, with weak spin-orbit coupling and long spin diffusion length, serves as an ideal spacer that maintains the spin pumping effect while decoupling the two magnetic layers.[47] If the spin pumping effect were the dominant mechanism, the magnetic damping should remain unaffected by inserting a thin Al layer. Fig. 3(a) compares the $\alpha$ of $Mn_2Au(2)/Al/CoFeB$ and $Mn_2Au(2)/CoFeB$. Inserting the Al interlayer reduces $\alpha$ below 160 K to about 0.012, a value identical to that of $Mn_2Au(2)/CoFeB$ above 160 K. The elimination of the damping increment by the Al interlayer indicates that the enhanced damping is caused not by spin pumping, but rather by an interaction requiring direct contact between $Mn_2Au$ and CoFeB layers. Given that SQUID measurements demonstrate the emergence of AFM order in 2-nm-thick $Mn_2Au$ below 40 K and the synchronous increase of magnetic damping and $H_{rot}$, we attribute the increase in $\alpha$ to the spin angular momentum transfer, mediated by exchange coupling between AFM $Mn_2Au$ and FM CoFeB. As the temperature decreases, the strengthened $Mn_2Au$ AFM order further enhances this exchange coupling and thus the spin angular momentum transfer, leading to the enhancement of magnetic damping.

As the temperature decreases from RT to 160 K, the $Mn_2Au$ layer remains paramagnetic. The absence of AFM order precludes exchange coupling between $Mn_2Au$ and CoFeB, resulting in negligible $H_{rot}$ acting on the CoFeB layer. Consequently, spin angular momentum transfer from CoFeB to $Mn_2Au$ is suppressed, leading to negligible damping variation.

As the temperature decreases from 160 K to $T_N$ (~40 K, as concluded from SQUID measurements), the emerging interfacial AFM-FM exchange coupling, established by short-range AFM order in $Mn_2Au$, enables spin angular momentum transfer and dissipation from CoFeB to $Mn_2Au$, thereby enhancing the magnetic damping. Key evidence supporting this mechanism is the synchronous increase in the exchange coupling field $H_{rot}$, a direct manifestation of interfacial exchange coupling between AFM spins in $Mn_2Au$ and FM spins in CoFeB.[36,37,48] Although macroscopic AFM order in $Mn_2Au$ only establishes below $T_N$, the short-range AFM order well above $T_N$ has been widely reported,[49–51] and these fluctuating AFM spins can generate $H_{rot}$ by exchange coupling with FM.[39,52,53] Meanwhile, this exchange coupling enables the precessing magnetization of CoFeB to exert a torque on $Mn_2Au$, thereby transferring the spin angular momentum to $Mn_2Au$.[20,54] The transferred spin angular momentum is efficiently relaxed due to the strong spin-orbit coupling in $Mn_2Au$, manifesting as spin angular momentum dissipation.[55] Consequently, with decreasing temperature, the short-range AFM order of $Mn_2Au$ strengthens, enhancing the AFM-FM exchange coupling and facilitating spin angular momentum transfer and dissipation from CoFeB into $Mn_2Au$, thus leading to the increase in $\alpha$. Fig. 4 schematically illustrates this transfer and dissipation process. As the temperature decreases below 160 K, AFM order progressively establishes in the $Mn_2Au$ layer, thereby enhancing the AFM-FM exchange coupling between $Mn_2Au$ and CoFeB. These effects collectively strengthen $H_{rot}$ and drive an efficient spin angular momentum transfer and dissipation, leading to the substantial enhancement of $\alpha$.

Below $T_N$, the establishment of macroscopic AFM order in $Mn_2Au$ further strengthens the exchange coupling, leading to further enhancement of both $H_{rot}$ and $\alpha$. The exchange coupling between the $Mn_2Au$ AFM order and the CoFeB magnetization directly generates $H_{rot}$, and simultaneously facilitates spin angular momentum transfer and dissipation, thereby accounting for the observed increase of damping $\alpha$. Several previous studies on AFM/FM bilayers have suggested that the AFM-FM exchange coupling is the possible mechanism behind magnetic damping,[20,21,23,56,57] but the key physical parameters characterizing this coupling, such as the exchange coupling field $H_{rot}$, remain insufficiently explored. Our study not only quantitatively determines the value of the exchange coupling field, but also, more importantly, reveals a synchronous trend between magnetic damping and the AFM-FM exchange coupling field.

We further investigated the temperature-dependent $\alpha$ and $H_{rot}$ in $Mn_2Au$/CoFeB with 4-nm-thick $Mn_2Au$ layer (see supplementary material, Fig. S6). $Mn_2Au$(4)/CoFeB exhibits a synchronized increase in $H_{rot}$ and $\alpha$ as temperature decreases from RT to 100 K. This observation can be understood by the thickness-dependent $T_N$ in AFM materials. Previous studies have demonstrated that critical temperature ($T_N$ or $T_C$) scales with film thickness.[22,28,29,50,58,59] Ultrathin AFM films exhibit a suppressed $T_N$, while $T_N$ of thicker films approaches the bulk value. Therefore, the increased $Mn_2Au$ thickness elevates $T_N$ toward the bulk value (>1000 K), causing the enhancement of $\alpha$ and $H_{rot}$ to emerge synchronously at around RT in $Mn_2Au$(4)/CoFeB. This feature contrasts sharply with that of $Mn_2Au$(2)/CoFeB, where the damping enhancement only appears below 160 K. This explicit $Mn_2Au$ thickness dependence provides compelling evidence that the synchronized enhancement of $H_{rot}$ and $\alpha$ is intrinsically linked to the AFM order of the

$Mn_2Au$ layer.

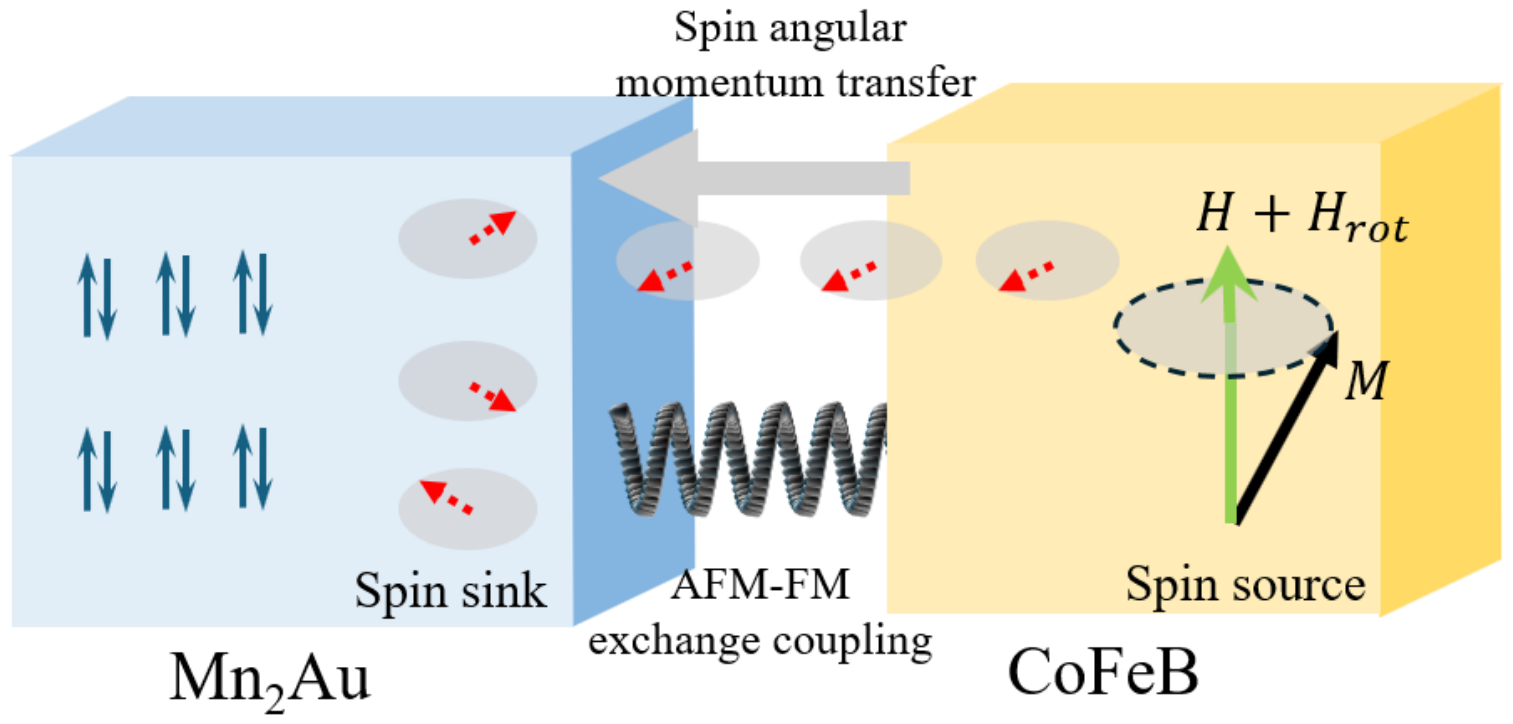

**FIG. 4.** Schematic illustration of the physical mechanism of spin angular momentum transfer and dissipation in the $Mn_2Au(2)/CoFeB$ bilayer. Below 160 K, AFM order (represented by paired blue arrows in the $Mn_2Au$ layer) and AFM-FM exchange coupling (depicted as the spring between $Mn_2Au$ and CoFeB) are established and progressively enhanced with decreasing temperature. This leads to a synchronized increase of $H_{rot}$ and efficient spin angular momentum transfer (illustrated by the gray span with the dashed red arrows moving from CoFeB into $Mn_2Au$) and dissipation (illustrated by the randomly arranged dashed red arrows in $Mn_2Au$), thereby enhancing the magnetic damping $\alpha$.

In summary, we investigate the temperature-dependent magnetic dynamics of $Mn_2Au$/CoFeB bilayers using FMR. For the $Mn_2Au$(2 nm)/CoFeB bilayer, the magnetic damping $\alpha$ remains unchanged and the rotatable anisotropic field $H_{rot}$ is negligible above 160 K. Below this temperature, both the parameters increase dramatically and synchronously. The damping enhancement is attributed to spin angular momentum transfer and dissipation mediated by the exchange coupling between $Mn_2Au$ and CoFeB, which is enhanced by $Mn_2Au$ AFM ordering as temperature decreases. Analogous enhancement of magnetic damping is also observed in the $Mn_2Au$(4 nm)/CoFeB bilayer near room temperature, conclusively confirming its dependence on the $Mn_2Au$ AFM ordering. These findings provide a deeper insight into magnetic dynamics in AFM/FM bilayers and have implications for the design and application of spintronic devices.

This work is supported by the National Key Research and Development Program of China (No. 2024YFA1408801), National Natural Science Foundation of China (Grant No. 62427901, 12574125, T2394473), and the Natural Science Foundation of Jiangsu Province of China (BK20211144).

## SUPPLEMENTARY MATERIAL

See the supplementary material for additional information on magnetic hysteresis loops of field-cooled $Mn_2Au(2)/CoFeB$ samples, in-plane angular-dependent resonant field, out-of-plane magnetic hysteresis loop for $Mn_2Au(2)/CoFeB$, *M*-*T* curve of $Mn_2Au(2)/CoFeB$ by SQUID, polar angle-dependent FMR linewidth, temperature-dependent magnetic damping and $H_{rot}$ of $Mn_2Au(4)/CoFeB$.

## AUTHOR DECLARATIONS

### Conflict of Interest

The authors have no conflicts to disclose.

## DATA AVAILABILITY

The data that support the findings of this study are available from the corresponding author upon reasonable request.

# Supplementary Material

## Antiferromagnetic Ordering Enhanced Magnetic Damping in $Mn_2Au$/CoFeB Bilayers

Donghang Xie[1,2] , Haozhe Wang[1,3] , Zhe Zhang[1,2] , Zishuang Li[1,3] , Jiahua Lu[1,2] , Ronghua Liu[1,3,a)] , Jun Du[1,3], Bo Liu[1,2] , Yu Yan[1,2] , Liang He[1,2] , Jing Wu[4] , Rong Zhang[2] , Bo Liu[5], Tiejun Zhou[5], Yongbing Xu[1,2,4,a)] , Xuezhong Ruan[1,2,a)]

**AFFLIATIONS**

*1 State Key Laboratory of Spintronics, Nanjing University, Suzhou, 215163, China*

*2 Jiangsu Provincial Key Laboratory of Advanced Photonic and Electronic Materials, School of Electronic Science and Engineering, Nanjing University, Nanjing 210093, China*

*3 School of Physics, Nanjing University, Nanjing 210093, China*

*4 York-Nanjing Joint Center (YNJC) for Spintronics and Nano Engineering, Department of Electronics and Physics, The University of York, York YO10 5DD, UK*

*5 State Key Laboratory for Spintronic Devices and Technologies, Hangzhou 311305, China*

[a)] Authors to whom correspondence should be addressed: rhliu@nju.edu.cn, ybxu@nju.edu, and xzruan@nju.edu.cn.

# Supplementary Material

## Magnetic hysteresis loops of $Mn_2Au$(2)/CoFeB(4) after field cooling

Here, we present the magnetic hysteresis loops of $Mn_2Au$(2)/CoFeB(4) after field cooling in Figure S1. The sample was field cooled from RT to 4 K under 5 kOe. The coercive fields are indicated in the figure.

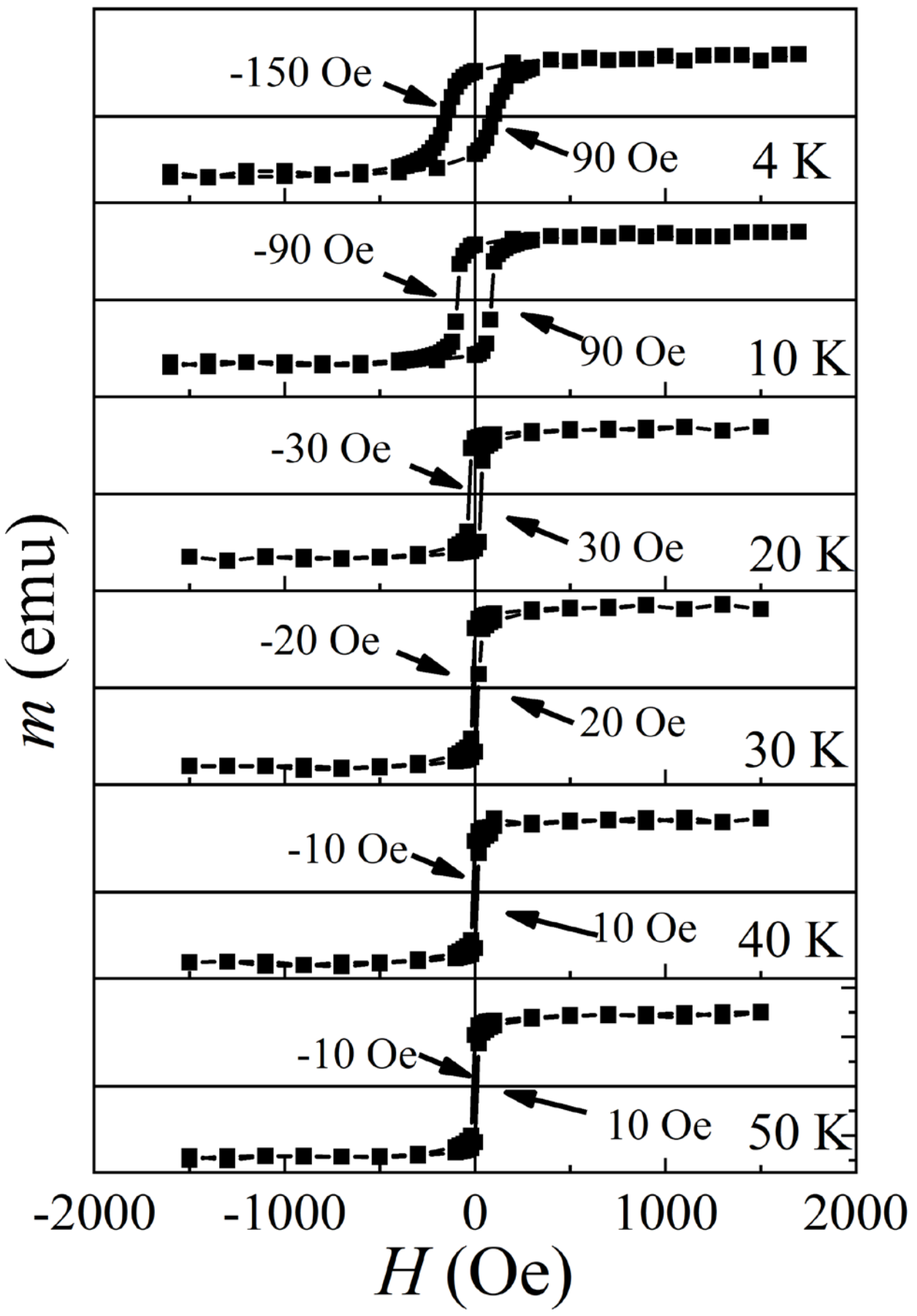

**Fig. S1** Magnetic hysteresis loops of field-cooled $Mn_2Au$(2 nm)/CoFeB(4 nm) measured at 4 K, 10 K, 20 K, 30 K, 40 K, and 50 K after cooling from room temperature under an in-plane field of 5 kOe. As temperature decreases, the loops exhibit exchange bias and enhanced coercivity, confirming macroscopic antiferromagnetic order in the 2-nm-thick $Mn_2Au$ layer at low temperature. The coercive fields are indicated in the figure.

## In-plane azimuthal-dependent $H_{res}$ at 13 GHz and 10 GHz

Here in Fig. S2, we present the in-plane azimuthal-dependent $H_{res}$ at 13 GHz and 10 GHz under 280 K, 200 K, 140 K, 90 K, and 40 K. $H_{res}$ exhibit in-plane isotropic behavior at all the measured temperatures.

# Supplementary Material

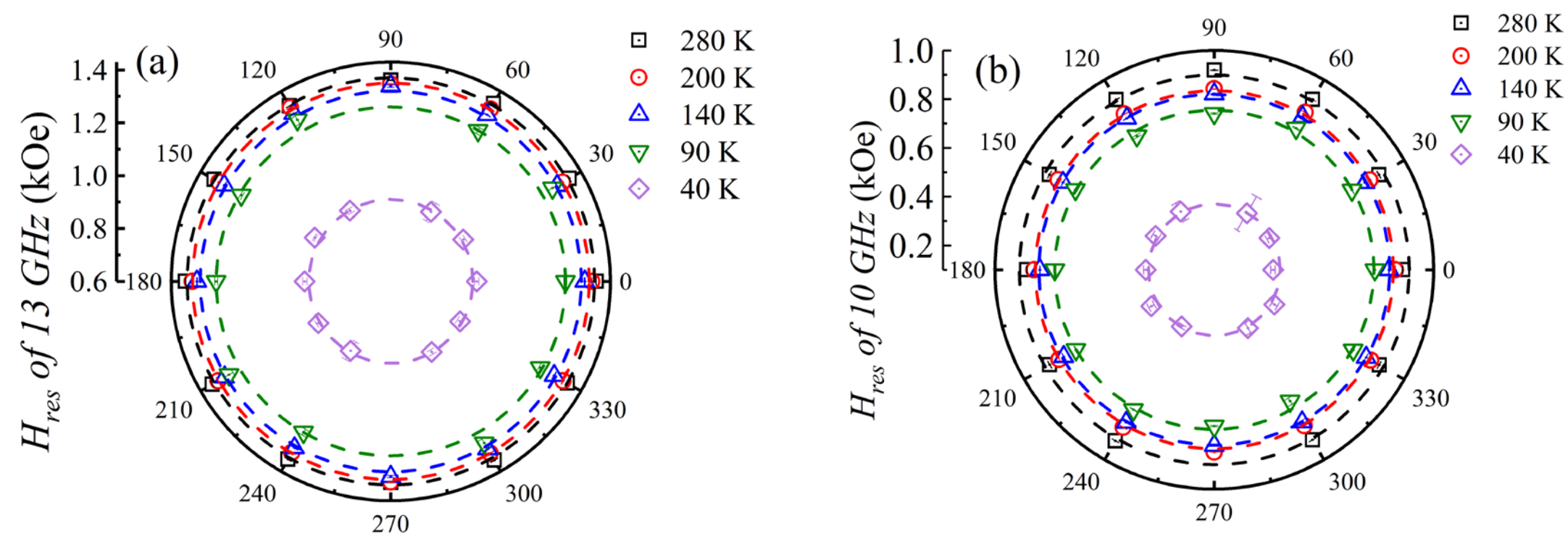

**Fig. S2** In-plane azimuthal-dependent resonance field ($H_{res}$) of $Mn_2Au$(2 nm)/CoFeB(4 nm) measured at 13 GHz (a) and 10 GHz (b), at temperatures of 280 K, 200 K, 140 K, 90 K, and 40 K. $H_{res}$ exhibits isotropic behavior at all measured temperatures, indicating that the anisotropy field is independent of in-plane direction. In Fig. S2(a), the raw data were measured only from -90°-90°, while the 90°-270° data were replicated from the -90°-90° data by symmetry.

**Out-of-plane (OOP) magnetic hysteresis loop of $Mn_2Au$(2)/CoFeB(4) at room temperature.** The OOP loop is shown in Figure S3. The saturation field is 14.5 kOe as indicated in the figure.

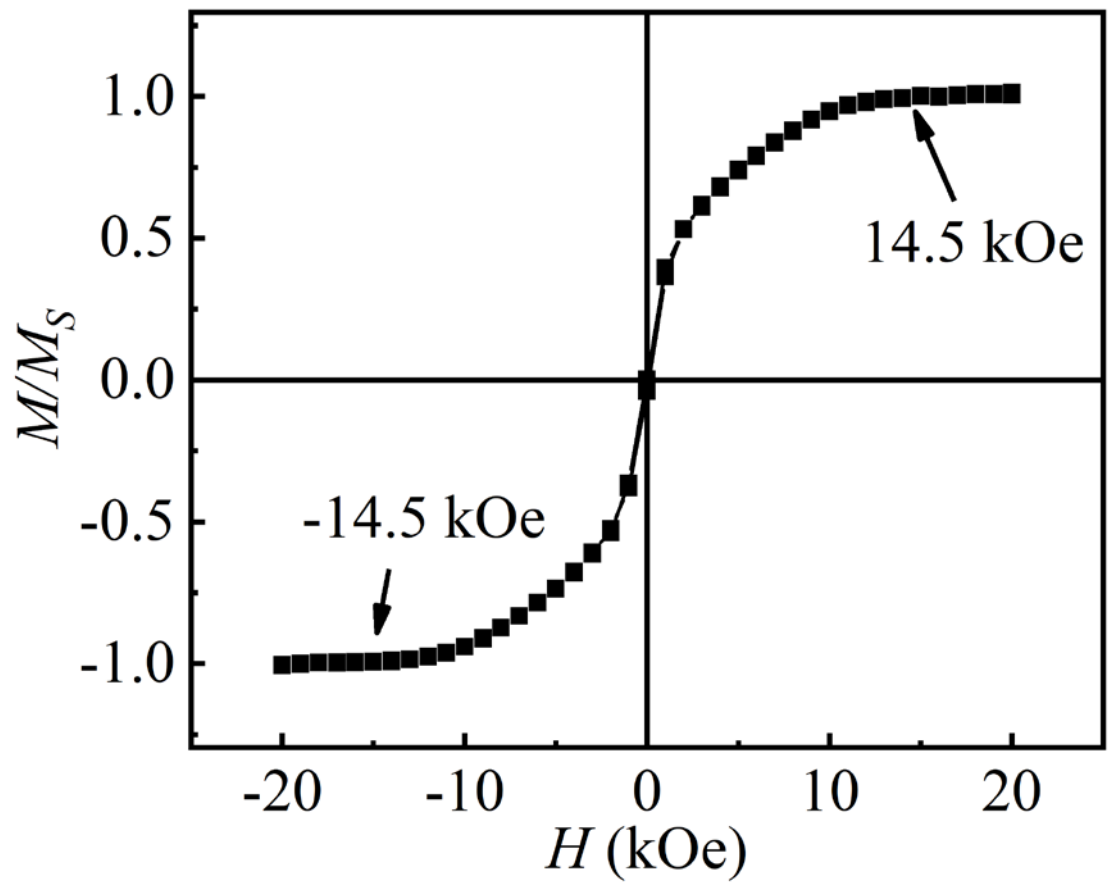

**Fig. S3** Out-of-plane hysteresis loop of $Mn_2Au$(2)/CoFeB(4) at room temperature. The saturation field is 14.5 kOe as indicated in the figure.

**Field-Cooling Magnetization-Temperature (*M-T*) Curve of $Mn_2Au$(2)/CoFeB(4) by Superconducting Quantum Interference Device (SQUID)**

# Supplementary Material

The *M-T* curve of field cooling $Mn_2Au(2)/CoFeB(4)$ is shown in Figure S4, which was measured from 300 K to 4 K under a 5 kOe magnetic field applied in the film plane.

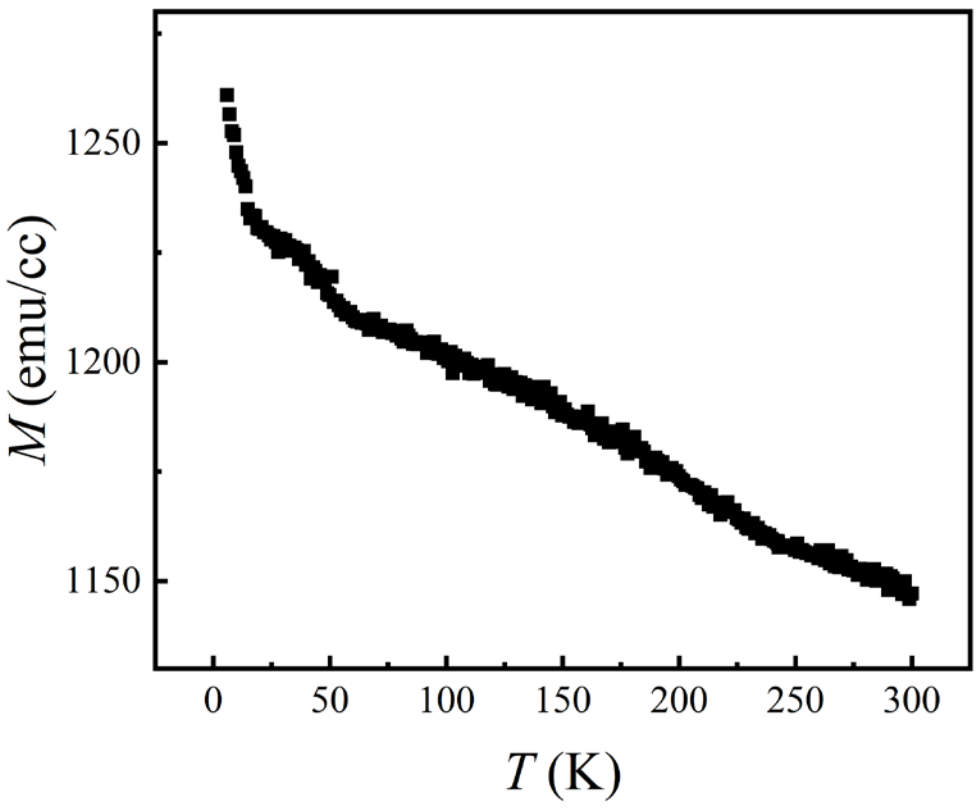

**Fig. S4** Temperature-dependent saturation magnetization $M_S$ of $Mn_2Au(2)/CoFeB(4)$ measured during field cooling from 300 K to 4 K under a 5 kOe in-plane magnetic field. The $M_S$ increases monotonically as temperature decreases, consistent with the typical behavior of CoFeB.

**Polar angle ($\theta_H$)-dependent FMR linewidth in $Mn_2Au(2)/CoFeB(4)$**

To further rule out two-magnon scattering as the possible mechanism, we carried out the polar angle ($\theta_H$) -dependent FMR linewidth in $Mn_2Au(2)/CoFeB(4)$ at temperatures of 80 K, 100 K, 120 K, and 140 K, where AFM order-induced damping enhancement emerges. Fig. S5 shows the extracted $\theta_H$-dependent FMR linewidth. The signature of the TMS is characterized by a sharp drop in the FMR linewidth when $\theta_H$ approaches 0°, specifically after $\theta_H$ decreases below 10° during the rotation from 90° (in-plane) to 0° (out-of-plane). However, in Fig. S5, we observed that in our $Mn_2Au/CoFeB$ bilayer, as the external applied field rotates from in-plane to out-of-plane, the linewidth increase monotonically. This behavior is in stark contrast to that of the TMS. This increase as the field direction rotates from in-plane to out-of-plane is commonly attributed to the formation of multidomain structures due to fluctuation in the demagnetizing field[1]. In our system, the linewidth monotonically increases as $\theta_H$ decreases from 90° to 0°, indicating that TMS is not the dominant mechanism.

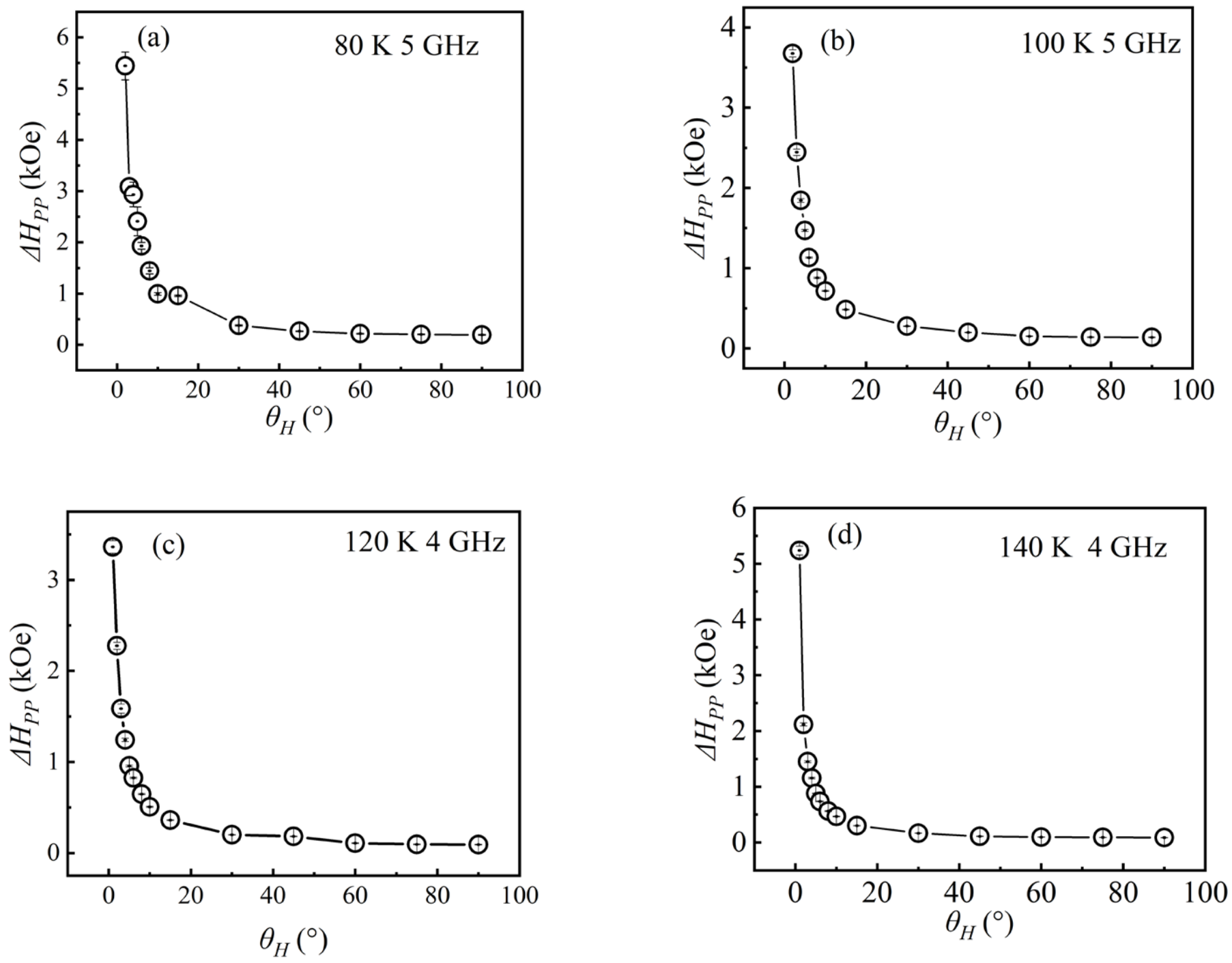

**Fig. S5** Polar angle ($\theta_H$) -dependent FMR linewidth in $Mn_2Au(2)/CoFeB(4)$ at (a) 80 K, 5 GHz, (b) 100 K, 5 GHz, (c) 120 K, 4 GHz, (d) 140 K, 4 GHz. The linewidth increases monotonically as the external field rotates from in-plane to out-of-plane, without a sharp drop near 0° that is characteristic of two-magnon scattering. This behavior rules out two-magnon scattering as the dominant mechanism responsible for the enhanced damping in $Mn_2Au/CoFeB$.

**Temperature-dependent Magnetic Gilbert damping and $H_{rot}$ of $Mn_2Au(4)/CoFeB(4)$**

Here, we present the temperature-dependent magnetic Gilbert damping $\alpha$ and rotatable anisotropic field $H_{rot}$ in $Mn_2Au(4)/CoFeB(4)$. The $\alpha$ and $H_{rot}$ exhibit synchronous increase by decreasing temperature from room temperature 120 K, in stark contrast to those of $Mn_2Au(2)/CoFeB(4)$, in which the enhancement only appears below 160 K. This explicit $Mn_2Au$ thickness dependence provides compelling evidence confirming that the synchronized enhancement of $H_{rot}$ and $\alpha$ is essentially linked to the AFM ordering of the $Mn_2Au$ layer, with the critical temperature for this effect strongly depending on the $Mn_2Au$ thickness.

# Supplementary Material

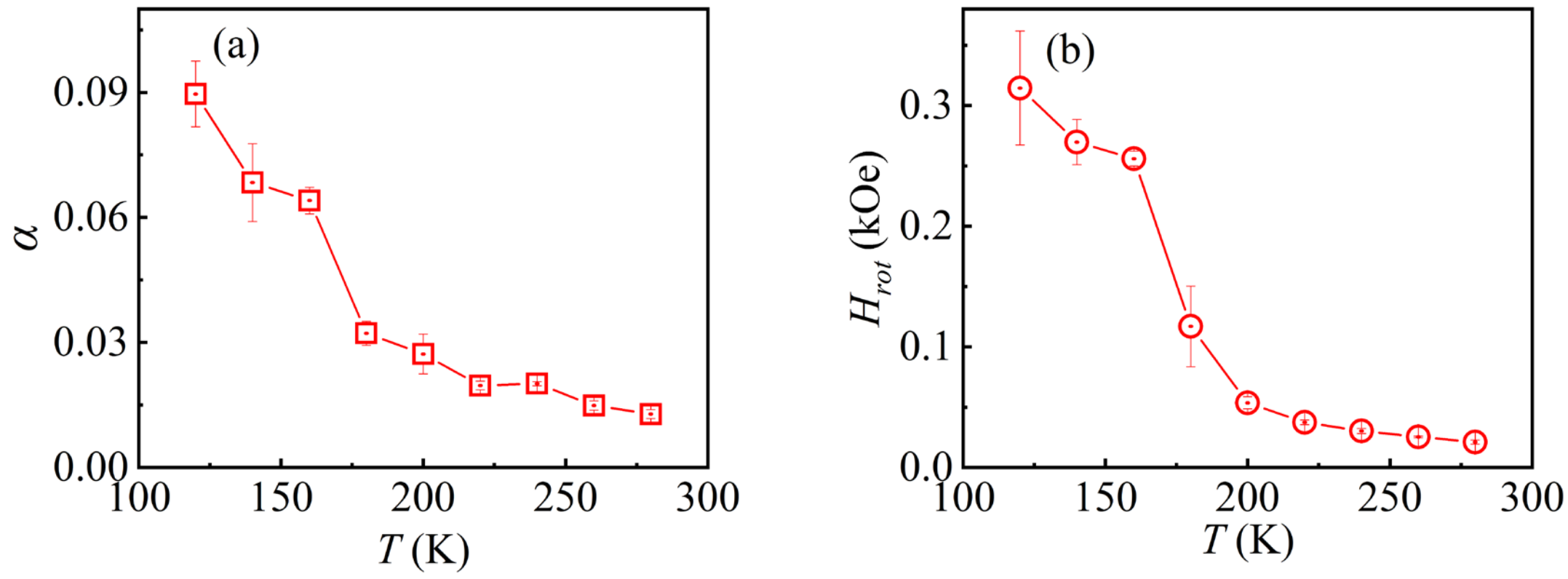

**Fig. S6** Temperature-dependent (a) magnetic Gilbert damping $\alpha$ and (b) rotatable anisotropic field $H_{rot}$ in $Mn_2Au$(4)/CoFeB(4). Both these parameters exhibit a synchronous increase as the temperature decreases from room temperature to 120 K.